\documentclass[aps,floatfix,showpacs,twocolumn]{revtex4}
\usepackage{amsmath}
\usepackage{amsfonts}
\usepackage{amssymb}
\usepackage{graphics,epsfig}
\usepackage{graphicx}
\begin{document}

\title{Some notes on the Kruskal - Szekeres completion.}
\author{Kayll Lake \cite{email}}
\affiliation{Department of Physics, Queen's University, Kingston,
Ontario, Canada, K7L 3N6 }
\date{\today}
\begin{abstract}
The Kruskal - Szekeres (KS) completion of the Schwarzschild spacetime is open to Synge's methodological criticism that the KS procedure generates ``good" coordinates from ``bad". This is addressed here in two ways: First I generate the KS coordinates from Israel coordinates, which are also ``good", and then I generate the KS coordinates directly from a streamlined integration of the Einstein equations.
\end{abstract}
\pacs{04.20.Cv, 04.20.Jb, 04.20.Gz}

\maketitle

There can hardly be any modern introduction to general relativity that does not discuss the Kruskal \cite{Kruskal} - Szekeres \cite{Szekeres} (KS) completion of the Schwarzschild spacetime. Both these papers make reference to an earlier work by Synge \cite{Synge}. Whereas the Kruskal reference regards only a reference by Synge to a lecture given by H. P. Robertson  in Toronto in 1939 (regarding the singular structure of the Schwarzschild spacetime), Szekeres states that the now famous coordinate transformations ``are essentially due to Synge". Indeed, after considerable labor,  Synge arrives at a complete understanding of what has become known as the ``Kruskal diagram" (see his Figure 8). Years latter, and curiously without reference to his earlier work, Synge \cite{Synge1} viewed the KS procedure as open to methodological criticism in that defective coordinates where used to generate regular coordinates. This lead him in \cite{Synge1} to develop KS coordinates directly from Einstein's equations. Whereas the KS procedure is standard in most modern texts, a few do follow Synge's direct approach \cite{chandra}. In this pedagogical note, heeding Synge's criticism, I generate KS coordinates from Israel coordinates and I also generate the KS coordinates directly from a streamlined integration of the Einstein equations.
\bigskip

As shown elsewhere \cite{lake}, a direct integration of the Einstein equations gives the following complete covering of the Schwarzschild manifold:
\begin{equation}\label{israelmetric}
ds^2=\frac{2w^2}{8M^2-uw}du^2-2dudw+(2M-\frac{uw}{4M})^2d\Omega^2_{2},
\end{equation}
where $M$ is a constant and $d\Omega^2_{2}$ is the metric of a unit two-sphere ($d\theta^2+\sin^2 \theta d \phi^2$). This form was first given by Israel \cite{israel} by way of coordinate transformations. The coordinates $u$ and $w$ in (\ref{israelmetric}) have
dimensions of length. It is convenient to introduce dimensionless
coordinates $\tilde{u}$ and $\tilde{w}$ defined as follows:
\begin{equation}\label{dimensionless}
u \equiv M \tilde{u}, \;\;\;\; w \equiv 8M \tilde{w}
\end{equation}
where $M>0$ and the number $8$ is introduced purely as a matter of
convenience. Under the transformations (\ref{dimensionless}) we
obtain
\begin{equation}\label{basemetric}
ds^2=(2M)^2 d\tilde{s}^2
\end{equation}
where
\begin{equation}\label{basemetricform}
d\tilde{s}^2=\frac{4w^2}{1-uw}du^2-4dudw+(1-uw)^2d\Omega^2_2
\end{equation}
and we have dropped the $\tilde{~}$ on $u$ and $w$ in
(\ref{basemetricform}). Trajectories
with tangents $k^{\alpha}=\delta^{\alpha}_{w}$ (constant $u, \theta$ and $\phi$) are radial null geodesics affinely parameterized by $w$, which we take increasing to the future.  Trajectories associated with the vector field $l^{\alpha}=\delta^{\alpha}_{u}$
(constant $w, \; \theta$ and $\phi$, and we take $u$ also increasing along the associated trajectory) are radial
null geodesics (affinely parameterized by $u$) only for $w=0$. More generally, for $w \neq 0$,
\begin{equation}\label{newgeodesic}
m^{\alpha}\equiv (\frac{1-uw}{w},w,0,0)
\end{equation}
satisfies $m^{\alpha}m_{\alpha}=m^{\beta}\nabla_{\beta}m^{\alpha}=0$.
The associated null geodesics can be written as
\begin{equation}\label{mgeodesics}
uw=\ln(\frac{w}{\delta})
\end{equation}
where $\delta$ is a constant.

\bigskip

The null geodesics given by (\ref{mgeodesics}) can be
rewritten as
\begin{equation}\label{firstlambert}
w=-\frac{\mathcal{L}(-u  \delta)}{u}
\end{equation}
where $\mathcal{L}$ is the Lambert W function \cite{lambert}. Since $\delta$ labels individual radial null geodesics, relation
(\ref{firstlambert}) suggests the introduction of the null
coordinate $v$ defined by \cite{scale}
\begin{equation}\label{nullv}
w=-\frac{\mathcal{L}}{u},
\end{equation}
where, and henceforth,
\begin{equation}\label{L}
    \mathcal{L} \equiv \mathcal{L}(-\frac{u v}{e}).
\end{equation}
The resultant coordinate transformation is
\begin{equation}\label{finalv}
v=we^{1-uw},
\end{equation}
which is well defined at $w=0$.
Under this transformation the metric (\ref{basemetricform}) takes
the form
\begin{equation}\label{ks2}
d\tilde{s}^2=\frac{-4}{(1+\mathcal{L})e^{1+\mathcal{L}}}dudv+(1+\mathcal{L})^2d\Omega^2_{2}
\end{equation}
where (\ref{basemetric}) still holds. This is the Kruskal - Szekeres representation. Trajectories
with tangents $\mathcal{K}^{\alpha}=e^{\mathcal{L}}(1+\mathcal{L}) \delta^{\alpha}_{v}$ (constant $u=u_{0}, \theta$ and $\phi$) are radial null geodesics given by
\begin{equation}\label{u0}
    v(\lambda)=\lambda e^{-\frac{u_{0} \lambda}{e}}
\end{equation}
where $\lambda$ is an affine parameter (defined, of course, only up to a linear transformation) and we note the expansion
\begin{equation}\label{expK}
    \nabla_{\alpha}\mathcal{K}^{\alpha}=\frac{-2u_{0}}{e(1+\mathcal{L})}.
\end{equation}
Trajectories
with tangents $\mathcal{M}^{\alpha}=e^{\mathcal{L}}(1+\mathcal{L}) \delta^{\alpha}_{u}$ (constant $v=v_{0}, \theta$ and $\phi$) are radial null geodesics given by
\begin{equation}\label{v0}
    u(\lambda)=\lambda e^{-\frac{v_{0} \lambda}{e}}
\end{equation}
and we now note the expansion
\begin{equation}\label{expm}
    \nabla_{\alpha}\mathcal{M}^{\alpha}=\frac{-2v_{0}}{e(1+\mathcal{L})}.
\end{equation}
On the horizons $u=0$ and $v=0$ then $v$ and $u$ are affine parameters. The only singularity in (\ref{ks2}) occurs for $\mathcal{L}=-1$. That is, $uv=1$.

\bigskip

Let us now proceed more directly, streamlining the procedure pioneered by Synge \cite{Synge1}. Let us write
\begin{equation}\label{newks}
    ds^2=-f du dv+r^2\Omega^2_{2},
\end{equation}
where $f$ and $r$ are functions of $u$ and $v$. The vanishing of the Ricci tensor requires \cite{symmetric}
\begin{equation}\label{ricci1}
    \frac{\partial r}{\partial v}=-B(1-\frac{2M}{r})
\end{equation}
and
\begin{equation}\label{ricci2}
    f=2B\frac{\partial r}{\partial u}
\end{equation}
where $B$ is an arbitrary function of $v$ and again $M$ is a constant. The solution to (\ref{ricci1}) is
\begin{equation}\label{soln}
    r(u,v)=2M(\mathcal{L}(\psi)+1)
\end{equation}
where
\begin{equation}\label{psi}
    \psi=-\frac{\exp(-\frac{1}{2M}(\int B dv+C+2M))}{2M}
\end{equation}
and $C$ is an arbitrary function of $u$. To achieve the range $0<r<\infty$ the functions $B$ and $C$ are not arbitrary. Rather, up to disposable constants, the unique choices are
\begin{equation}\label{uniqueb}
    B=-\frac{2M}{v}
\end{equation}
and
\begin{equation}\label{uniquec}
    C=-2M\ln(2Mu).
\end{equation}
With (\ref{uniqueb}) and (\ref{uniquec}), (\ref{newks}) reduces to (\ref{basemetric}) with $d\tilde{s}^2$ given by (\ref{ks2}).
\begin{acknowledgments}
This work was supported by a grant from the Natural Sciences and
Engineering Research Council of Canada. Portions of this work were
made possible by use of \textit{GRTensorII} \cite{grt}.
\end{acknowledgments}


\begin{thebibliography}{}
\bibitem{scale} Our choice of scale for $v$, though somewhat inconvenient, reflects common usage.
\bibitem{symmetric}The metric (\ref{newks}) is, of course, symmetric in $u$ and $v$. The symmetric counterparts to (\ref{ricci1}) and (\ref{ricci2}) are not needed.
\end{thebibliography}
\end{document}